\documentstyle[12pt]{article}

\newcommand{\be}{\begin{equation}}
\newcommand{\ee}{\end{equation}}
\newcommand{\bea}{\begin{eqnarray}}
\newcommand{\eea}{\end{eqnarray}}
\newcommand{\ba}[1]{\begin{array}{#1}}
\newcommand{\ea}{\end{array}}
\newcommand{\noi}{\noindent}

\newcommand{\lsim}{\mathrel{\lower4pt\hbox{$\sim$}}
\hskip-12.5pt\raise1.6pt\hbox{$<$}\;}

\newcommand{\gsim}{\mathrel{\lower4pt\hbox{$\sim$}}
\hskip-12.5pt\raise1.6pt\hbox{$>$}\;}

\begin{document}

\centerline{\bf Extra Long Baseline Neutrino Oscillations and CP
Violation}
\vspace{.2in}

\centerline{William J. Marciano}
\centerline{\it Brookhaven National Laboratory, Upton, NY\ \ 11973}
\vspace{.5in}

\centerline{\bf Abstract}
\begin{quote}

The potential for studying CP violation in neutrino oscillations using
conventional $\nu_\mu$ and $\bar\nu_\mu$ beams is examined. For $\Delta
m^2_{21}<< \Delta m^2_{31}$ and fixed neutrino energy, $E_\nu$, the CP
violating asymmetry $A\equiv P(\nu_\mu\to\nu_e)-P(\bar\nu_\mu\to\bar\nu_e)
/P(\nu_\mu\to\nu_e)+P(\bar\nu_\mu \to\bar\nu_e)$ in vacuum is shown to be
measurable with roughly equal maximal statistical precision at distances
$L_n\simeq (2n+1) \left(\frac{2\pi E_\nu}{\Delta m^2_{31}}\right),
n=0,1,2\dots$ (up to some $n$ in the leading $\Delta m^2_{21}$
approximation). For extra long baselines, $n\ge1$, the falloff in
detected oscillation events, $N=N_{\nu_e}+N_{\bar\nu_e}$, by $\sim1/(2n+1)^2$
is compensated by a factor $\sim(2n+1)$ increase in the asymmetry such
that the statistical figure of merit F.O.M.${}\equiv (\delta A/A)^{-2}
= A^2N/1-A^2$ is approximately independent of $n$. However, for the
larger $n\ge1$ asymmetries, some backgrounds as well as systematic
uncertainties from flux normalization, detector acceptance etc.\ which
scale with distance as $1/(2n+1)^2$ are shown to be relatively
suppressed in $\delta A/A$ by $\sim 1/(2n+1)$. Also, low energy
$E_\nu\simeq {\cal O}(1$GeV) extra long baseline experiments with
$L_n\simeq 1200$--2900 km (corresponding to $\Delta m^2_{31}\simeq
3\times 10^{-3}$ eV$^2$ and $n=1$--3) are better matched to the remote
locations of some proposed very large proton decay detectors which
would be essential for neutrino CP violation studies. Effects of matter
and realistic neutrino beam energy spread on  extra long baseline CP
violation experiments are briefly discussed.
\end{quote}

Experimental evidence for neutrino masses, mixing and oscillations has
become compelling. Atmospheric neutrino flux measurements by the
Super-Kamiokande (SK) collaboration \cite{kamio} indicate large mixing
$\nu_\mu\to\nu_\tau$ oscillations governed by a neutrino mass squared
difference

\be
\Delta m^2_{32} \equiv m^2_3 - m^2_2 \simeq (1.6-4)\times10^{-3} {\rm
eV}^2 \label{eq1}
\ee

\noi (The sign of $\Delta m^2_{32}$ is not yet determined, but here we
assume $m_3>m_2$.) Also, recent
results \cite{ahmad} from SNO on the solar $\nu_e$ flux when combined
with earlier SK and other \cite{cleve} solar neutrino data strongly
point to the so-called Large Mixing Angle (LMA) scenario with
\cite{bahcall} 

\be
\Delta m^2_{21}\equiv m^2_2-m^2_1\simeq (2-10)\times10^{-5}{\rm eV}^2
\label{eq2} 
\ee

\noi In terms of the $3\times3$ mixing matrix

\be
\left(
\ba{c} \nu_e \\ \nu_\mu \\ \nu_\tau \ea \right)
= \left( \ba{ccc} c_1c_3 & s_1c_3 & s_3 e^{-i\delta} \\
-s_1c_2-c_1s_2s_3e^{i\delta} & c_1c_2-s_1s_2s_3e^{i\delta} & s_2c_3 \\
s_1s_2-c_1c_2s_3e^{i\delta} & -c_1s_2-s_1c_2s_3e^{i\delta} & c_2c_3 
\ea \right)
\left(\ba{c} \nu_1 \\ \nu_2\\ \nu_3 \ea \right) \label{eq3}
\ee

\[
c_i\equiv\cos\theta_i\quad ,\quad s_i\equiv \sin\theta_i \quad , \quad
i=1,2,3 
\]

\noi those combined results imply relatively large mixing \cite{itow} 

\bea
\sin\theta_1 & \simeq & 0.4 \label{eq4} \\
\sin\theta_2 & \simeq & 0.707 \nonumber 
\eea

\noi On the other hand, failure to observe oscillations in
$\bar\nu_e\to\bar\nu_x$ reactor searches gives the bound \cite{chooz}
(for $\Delta m^2_{31} \simeq 3\times10^{-3}$ eV$^2$)

\be
\sin^22\theta_3 < 0.12 \qquad {\rm or} \qquad s_3<0.18 \qquad  (90\%
{\rm ~CL}) \label{eq5} 
\ee

\noi suggesting small first-third generation mixing. In the case of the
phase, $\delta$, nothing is currently known. Determination of $\delta$
and $\theta_3$ will tell us the amount of CP violation inherent to
lepton charged current mixing via the Jarlskog CP violation invariant
\cite{jarls} 

\be
J_{CP} \equiv s_1s_2s_3c_1c_2c^2_3\sin\delta \label{eq6}
\ee

Future experimental efforts will aim to measure the above parameters
with high precision and search for potential deviations from
expectations due to ``new physics'' such as sterile neutrino mixing.
Towards those ends, the KamLAND reactor experiment \cite{piep} should
pin down the values of both $\Delta m^2_{21}$ and $\sin\theta_1$, if
the LMA solar neutrino solution is indeed correct. Similarly, $K2K$
\cite{ahn} and MINOS \cite{minos} should better determine $\Delta
m^2_{32}$ and $\sin\theta_2$ via $\nu_\mu$ disappearance. Further down
the road, a recent proposal \cite{itow} for a JHF-Kamioka
$\nu_\mu\to\nu_e$ oscillation experiment using a conventional horn
generated pion decay neutrino beam of $\sim 0.7$ GeV from their future
0.77 megawatt proton synchrotron and the large
(22kton) SK detector located 295 km away would search for
$\sin^22\theta_3$
down to 0.006 (a factor of 20 improvement over the current bound) and
measure $\Delta m^2_{32}$ to $\pm10^{-4}$ eV$^2$ and $\sin^22\theta_2$
to about $\pm1\%$.

The reach of the JHF-Kamioka proposal \cite{itow} illustrates the
significant potential of intense low energy conventional neutrino beams
for studying oscillations when a very large detector such as SK is
available. The next generation of proton decay detectors will have
fiducial volumes 20--40 times larger than SK and will probably provide our first
opportunity to probe CP violation in neutrino oscillations and
thereby determine the phase $\delta$. Indeed, the JHF-Kamioka neutrino
project \cite{itow} envisions a phase II upgrade of its accelerator
complex to 4 Megawatts, which when combined with a possible 1000 kton
Hyper-Kamiokande detector, could increase their yearly statistics by a
factor of 200. Studies \cite{itow} indicate that with such a facility, in 2 years
of $\nu_\mu$ and 6 years of $\bar\nu_\mu$ running, they would
statistically probe CP violation well below  $\delta 
\simeq \pi/9$, assuming systematic uncertainties in the comparison of
$\nu_\mu\to\nu_e$ and $\bar\nu_\mu\to\bar\nu_e$ and backgrounds can be controlled.

In this paper, we examine potential advantages of studying CP violation
with conventional $\nu_\mu$ and $\bar\nu_\mu$ beams at extra long
baseline distances

\[
\qquad\qquad\qquad\qquad L_n\simeq (2n+1) L_0 \quad , \quad n=1,2,3\dots
\qquad\qquad\qquad\qquad (7a) 
\]

\noi which are larger by odd multiples 3,5,7 etc.\  than the generally
considered first oscillation maximum

\[
\qquad\qquad L_0=\left(\frac{2\pi E_\nu}{\Delta m^2_{31}} \right)
\simeq 413 (E_\nu/{\rm GeV}) \left(\frac{3\times10^{-3}{\rm
eV}^2}{\Delta m^2_{31}} \right) km \qquad\qquad ~~(7b)
\]

\addtocounter{equation}{1}

\noi As we will show, to leading order in $\Delta m^2_{21} / \Delta
m^2_{31} <<1$, the statistical figure of merit (F.O.M.)

\be 
F.O.M. \equiv (\delta A/A)^{-2} \label{eq8}
\ee

\noi which describes ones ability to measure  $A$, $J_{CP}$ and
$\sin\delta$ is roughly independent of $n$ (up to some $n$ where the leading
$\Delta m^2_{21}$ approximation starts to fail). That finding implies
that CP violation experiments at the first few $L_n, n=0,1,2\dots$, are (roughly)
statistically equivalent to first approximation (until one runs out of
events). However, the extra 
long baselines with $n\ge1$ have a potentially significant advantage
when it comes to backgrounds and systematic uncertainties from neutrino
flux normalization, detector acceptance etc.\ which scale down by
$1/(2n+1)^2$ with the larger distances. Such effects are relatively
suppressed in $\delta A/A$ by $\sim 1/2n+1$. Furthermore, the longer
distances may be better matched to the remote locations of very large
proton decay detectors \cite{jung} which are unlikely to be located
$\lsim413 (E_\nu$/GeV$) \left(\frac{3\times10^{-3}{\rm eV}^2}{\Delta
m^2_{31}} \right)$ km from existing intense sources of $E_\nu\sim{\cal
O}$(1 GeV) neutrino beams such as the AGS at Brookhaven or the Fermilab
Booster. For example Brookhaven-Homestake${}=2540$km
Fermilab-Homestake${}=1290$km while Brookhaven-Carlsbad${}=2920$km and
Fermilab-Carlsbad${}=1770$km \cite{barger}. (An exceptional case is the
JHF-Hyper-Kamioka proposal \cite{itow} which is fixed at $\sim295$ km
and thus optimized for and constrained to $E_\nu \simeq 0.7
\left(\frac{3\times 10^{-3}{\rm eV}^2}{\Delta m^2_{31}} \right)$ GeV CP
violation studies.)

We begin by giving the exact $\nu_\mu\to \nu_e$ oscillation probability
in vacuum at a distance $L$ for neutrino energy $E_\nu$ (matter effects
are subsequently considered). In terms of the mixing angles in
(\ref{eq3}), one finds (for $\bar\nu_\mu\to\bar\nu_e$ and
$\nu_e\to\nu_\mu$, $J_{CP}\to -J_{CP}$)

\bea
P(\nu_\mu\to\nu_e) & = & 4(s^2_2s^2_3c^2_3 +J_{CP}\sin\Delta_{21})
\sin^2\frac{\Delta_{31}}{2} \nonumber \\
& & +2(s_1s_2s_3c_1c_2c^2_3 \cos\delta -s^2_1s^2_2s^2_3c^2_3) \sin
\Delta_{31} \sin \Delta_{21} \label{eq9} \\
& & +4(s^2_1c^2_1c^2_2c^2_3 +s^4_1s^2_2s^2_3c^2_3 -2s^3_1s_2s_3c_1c_2c^2_3
\cos\delta -J_{CP} \sin\Delta_{31}) \sin^2\frac{\Delta_{21}}{2}
\nonumber \\
& & +8(s_1s_2s_3c_1c_2c^2_3 \cos\delta - s^2_1s^2_2s^2_3c^2_3) \sin^2
\frac{\Delta_{31}}{2} \sin^2 \frac{\Delta_{21}}{2} \nonumber
\eea

\noi where $J_{CP}$ is given in (\ref{eq6}) and

\bea
\Delta_{31} & \equiv & \Delta m^2_{31} L /2 E_\nu \nonumber \\
\Delta_{21} & \equiv & \Delta m^2_{21} L / 2 E_\nu \label{eq10}
\eea

For studies of CP violation, one considers the asymmetry

\be
A\equiv \frac{P(\nu_\mu\to\nu_e) -P(\bar\nu_\mu\to\bar\nu_e)}{P
(\nu_\mu\to\nu_e) + P(\bar\nu_\mu\to\bar\nu_e)} \label{eq11}
\ee

\noi To leading order in $\Delta_{21}$ (assumed to be small), one finds

\bea
\qquad P(\nu_\mu\to\nu_e) \simeq 4s^2_2s^2_3c^2_3 \sin^2
\frac{\Delta_{31}}{2} + {\cal O}(\Delta_{21}) \qquad\qquad\qquad\qquad\quad
(12a) \nonumber  \\
\qquad A\simeq \frac{J_{CP}\sin\Delta_{21}}{s^2_2s^2_3c^2_3} \simeq
\frac{2 s_1c_1c_2\sin\delta}{s_2s_3} \left(\frac{\Delta
m^2_{21}}{\Delta m^2_{31}}\right) \frac{\Delta m^2_{31}L}{4E_\nu} +
{\cal O}(\Delta^2_{21}) \qquad\quad (12b)\nonumber
\eea

\addtocounter{equation}{1}
\noi The asymmetry grows linearly with distance, but for fixed detector
size and neutrino energy, the flux of neutrinos
decreases as $\sim1/L^2$. The experimental determination of $A$ will
entail measuring $N_{\nu_e}$ events from $\nu_\mu\to\nu_e$ and
$N_{\bar\nu_e}$ from $\bar\nu_\mu\to\bar\nu_e$ oscillations such that

\be
A^{\rm exp} = \frac{N_{\nu_e}-rN_{\bar\nu_e}}{N_{\nu_e}+rN_{\bar\nu_e}}
\label{eq13} 
\ee

\noi where $r$ is the ratio of ($\nu_\mu$ flux)${}\times{}$($\nu_e$
cross-section)${}\times{}$($\nu_eN\to e^-N^\prime$
acceptance)${}\times{}$(running time) divided by the same quantity for
antineutrinos. The running times would normally (for small asymmetries)
be arranged to make $r=1$ up to systematic uncertainties. (For large
asymmetries, a different running strategy might be employed in which
run times are further relatively weighted by $\frac{1+A}{1-A}$.)

For our statistical discussion, we set $r=1$ and find the statistical figure of
merit (F.O.M.) regarding how well the asymmetry and therefore
$\sin\delta$ are measured is given by

\bea
F.O.M. & = & (\delta A/A)^{-2} = A^2 N/1-A^2 \nonumber \\
N & = & N_{\nu_e}+ N_{\bar\nu_e} \label{eq14}
\eea

\noi Note that it is linear in $N$ and quadratic in $A$. From
(12), one finds to leading order in $\Delta_{21}$ that the
statistical F.O.M. is optimized at the oscillation maxima

\be
\frac{\Delta_{31}}{2} = \frac{\Delta m^2_{31} L_n}{4E_\nu} = (2n+1)
\pi/2 \label{eq15}
\ee

\noi At distances $L_n$, the growth in $A$ by a factor of $(2n+1)$
compensates for the reduction in $N$ by $1/(2n+1)^2$ due to flux
reduction, leaving the F.O.M. roughly independent of $n$ (at least until
the small $\Delta_{21}$ approximation breaks down). Where our
approximation starts to break down depends on $s_3$ and $\Delta
m^2_{21}/\Delta m^2_{31}$ as our examples will subsequently illustrate.
Note also, that in our approximation the F.O.M. for $\delta A/A$ is
 insensitive to the value of $s_3$, at least until $s_3$ becomes so
small that the approximation in (12a) breaks down.

To illustrate the above finding and its domain of validity, we consider
measurements carried out at different $L_n$ for $E_\nu=1$ GeV beams
employing the following realistic parameters

\begin{center}
\begin{tabular}{cc}
$s_1 = 0.4$ & $c_1=0.916$ \\
$s_2 =0.707$ & $c_2=0.707$ \\
$s_3 =0.10$ & $c_3 = 0.995$ \\ 
$\sin\delta =0.707$ & $\cos\delta=0.707$ \\ \\
$\Delta m^2_{31}=3\times10^{-3}{\rm eV}^2$\qquad & \qquad$\Delta m^2_{21}
=5\times10^{-5} {\rm eV}^2$ 
\end{tabular}
\end{center}

\noi in the full expression of eq.~(\ref{eq9}). For definiteness, we assume running times (2
years of $\nu_\mu$ and 6 years of $\bar\nu_\mu$), fluxes and detector
assumptions given in phase II of the JHF-Hyper-Kamiokande proposal
\cite{itow} which envisions a 1000 kton water cerenkov detector and 4
megawatt proton synchrotron. We assume cuts on the events that reduce
the oscillation signals by about a factor of 2. In that way, our
statistics roughly represent simple scaling of the JHF-Hyper-Kamiokande
proposal. In an actual experiment
unconstrained by the 295 km requirement, one would probably employ somewhat
higher (average) energy neutrino beams, looser cuts and perhaps a wide band beam at larger distances.
However, those potential gains in statistics  might be somewhat offset
by oscillation 
losses at 
larger distances (see subsequent discussion); so, our table numbers
should be fairly realistic but are not to be taken too literally.

\begin{table}[ht]
\begin{center}
\begin{tabular}{lcccc}
$n$ & $L_n(km)$ & $N_{\nu_e}+N_{\bar\nu_e}$ & $A$ & F.O.M. \\
0 & 413 & 16000 & 0.134 & 293 \\
1 & 1239 & 1952 & 0.366 & 302 \\
2 & 2065 & 824 & 0.516 & 299 \\
3 & 2891 & 512 & 0.586 & 268 \\
4 & 3717 & 384 & 0.601 & 217 \\
5 & 4543 & 315 & 0.585 & 164 
\end{tabular}
\caption{Number of total oscillation events (after cuts)
$N_{\nu_e}+N_{\bar\nu_e}$, the CP violating asymmetry $A$, and F.O.M.
at different distances for the mixing parameters given above.
$E_\nu\simeq 1$ GeV, $r=1$  and the experimental 
conditions given in the text are  assumed. A
F.O.M${}=293$ corresponds to about a 17 sigma determination of
the asymmetry $A$.\label{tab1}} 
\end{center}
\end{table}

Table~\ref{tab1} nicely illustrates the main point of this paper. The
statistical F.O.M. is rather constant for $n=0$--3. Beyond those
distances, the $\Delta m^2_{21}$ governed oscillation starts to become
significant and eventually begins to reduce the asymmetry (although it remains quite
appreciable). Of course, at the larger distances, matter effects and
beam characteristics become more important and should be carefully
folded into any distance-beam energy optimization. 

To demonstrate the robustness of our result, we give in
table~\ref{tab2} changes brought about by reducing $s_3$ from 0.1 to
0.05. Note, that the oscillation statistics are reduced, but the F.O.M.
is rather unaffected (at least for $n=0$--2). Here the dilution due to
$\Delta_{21}$ oscillations  starts to set in earlier.

\begin{table}[ht]
\begin{center}
\begin{tabular}{lcccc}
$n$ & $L_n(km)$ & $N_{\nu_e}+N_{\bar\nu_e}$ & $A$ & F.O.M. \\
0 & 413 & 4168 & 0.257 & 295 \\
1 & 1239 & 616 & 0.580 & 312 \\
2 & 2065 & 328 & 0.650 & 240 \\
3 & 2891 & 248 & 0.600 & 140 \\
4 & 3717 & 216 & 0.536 & 87 
\end{tabular}
\caption{Parameters the same as table~\ref{tab1} except $s_3=0.05$
instead of 0.10.\label{tab2}} 
\end{center}
\end{table}

The measurement of CP violation as illustrated in tables \ref{tab1} and
\ref{tab2} is statistically significant, $\sim17$ sigma for the first
few $n$. One could
half the running 
time, descope the large water cerenkov detector to a more manageable
500kton, go from a 4 to a 1 Megawatt proton synchrotron
and still have a robust 4 sigma signal. However, a reduced asymmetry due to
a much smaller $\sin\delta \frac{\Delta m^2_{21}}{\Delta m^2_{31}}$
than the
0.012 value assumed here would
 require the higher statistics. One should have a better idea
regarding experimental demands as $\Delta m^2_{21}$, $\Delta m^2_{31}$
and the $\theta_i$ are more precisely determined.

Where the extra long baselines have a potential significant advantage
is in the area of systematic errors from flux normalization, detector
acceptance and those  backgrounds that scale with distance as
$1/(2n+1)^2$. To illustrate that point, we consider the experimental
asymmetry in (\ref{eq13}) where $r\simeq1$ contains uncertainties from
flux, cross-section, acceptance and other normalization effects. The
fractional systematic error $\delta r/r$ translates into the following
asymmetry error

\be 
\delta A/A \simeq \frac{1}{2} \; \frac{1-A^2}{A} \; \frac{\delta r}{r}
\label{eq16} 
\ee

\noi For larger asymmetries, the effect of $\delta r/r$ is
significantly reduced. For asymmetries exhibiting the $(2n+1)$ growth,
such systematics are reduced by $\sim 1/2n+1$. This advantage can be
significant, particularly if the CP violating asymmetry turns out to be
much smaller than
in our previous illustrative examples.

Other backgrounds, such as $\stackrel{(-)}{\nu}\!\!\!\!\!{}_e$ beam
contamination and misidentified neutral current residual events
remaining after cuts,  that scale down by $1/L^2$ will similarly be reduced
in relative importance by $\sim1/A$ for larger asymmetries. 
Non-distance dependent effects such as cosmic ray backgrounds would of
course be
worse for the reduced statistics  of extra long baseline
experiments. Fortunately, they are generally very small in deep underground
experiments  because of
directionality and the precise timing constraints imposed by beam
pulsing. Nevertheless, such effects are potentially more dangerous and
must be carefully studied.

For realistic experimental considerations, the full oscillation formula
in (\ref{eq9}) supplemented with matter effects \cite{wolf} must be
employed. In addition, neutrino beam energy spread, detector
acceptance, backgrounds etc.\ must be considered. Here, we make some
brief qualitative observations concerning such effects. They will be
further examined in a subsequent more detailed study.

At extra long distances, $L_n$, $n=1$--3, the smaller frequency
$\Delta_{21}$ oscillations and matter effects will effectively shift
somewhat the oscillation peaks. More important, matter effects will induce a
fairly significant CP non-violating (positive) asymmetry which must be
well controlled and understood in any serious study. For the distances
considered here $\sim1200$--2900 km and neutrino energies $E_\nu\sim
{\cal O}$(1 GeV), such calculations can be carried out in a
straightforward way that exactly includes $\Delta m^2_{31}$, $\Delta
m^2_{21}$ and matter density effects \cite{arafune}. Such a study has been nicely
illustrated in ref.~15. There, one finds for a distance of 2900 km
that matter effects are most important for the $n=0$ peak at $E_{\nu_\mu}\simeq 6$
GeV\null. At lower energies, $\nu_\mu$ oscillation peaks are also
somewhat shifted but less
amplified due to $\Delta m^2_{21}$ and matter effects. Those shifts may
influence the experimental beam energy setting for a given baseline
distance, but they  should not significantly change our discussion about
the statistical significance of extra long oscillation studies near the
first few $L_n$ with $n=1,2,3$ as long as $E_\nu$ is not too large. The
CP violating asymmetry increase at 
the longer distances should continue to compensate for the flux falloff in
the F.O.M. at least for the first few $n\ge1$. If the matter asymmetry
and CP violating asymmetry have the same sign (both positive), we can
have potentially enormous combined asymmetries at long distances with the
$1-A^2$ in the F.O.M. denominator playing a significant role.
Alternatively, for opposite signs, matter and CP violation asymmetries
will partially cancel. Of course, such a cancellation is also possible
and potentially more problematic for $n=0$. (In the JHF-Kamioka
discussion of ref.~5, a partial cancellation of matter and CP violating
asymmetries  which dilutes the signal is apparent for the phase chosen as an illustration
$\delta=-45^\circ$.) Generally, one expects better control of matter
induced asymmetries for the larger CP violating asymmetries of extra
long, $n\ge1$, baselines.

Also of importance for extra long beamline experiments is the neutrino
energy spectrum \cite{beavis}. For conventional pion decay generated
neutrino beams from $16\sim28$ GeV proton drivers,  one can either
employ a narrow band 
(energy spread) beam centered at some energy $E_\nu\sim1$ GeV by
placing the detector off-beam axis by some small angle or opt for a
more intense wide band beam with considerable support in the
$E_\nu\simeq 0.5$--2 GeV range \cite{itow,beavis}. Because the
oscillation peaks at $n\ge1$ have a narrower energy width, it will be
more challenging to tailor a narrow band beam to the detector distance
and oscillation parameters as $n$ grows. Instead, the wide band neutrino beam may
have advantages. In addition to the higher flux of neutrinos, the
energy spread would simultaneously cover several oscillation maxima.
For example, if the $E_\nu\simeq 2$ GeV component corresponds to the
$n=1$ peak (for $L_1\simeq2500$km), there will also be potentially
observable peaks at $E_\nu$
approximately 1.2 GeV ($n=2$), 
0.86 GeV ($n=3)$ and 0.67 GeV ($n=4$). Of course, there will also be
oscillation minima between the peaks that would reduce statistics
(roughly wiping out some flux gains of a wide band beam). Nevertheless,
an experiment that covers  and can partially unravel the $n=1$--4 oscillation peaks has the
potential (if statistics suffice) to not only probe CP violation, but
to determine the sign of $\Delta m^2_{31}$ and study the changing matter effects in
detail. Such potential capabilities provide strong motivation for extra
long baseline experiments. Those features will be explored in detail in
a subsequent study.

\end{document}